# COMPOSITIONAL TESTING FOR FSM-BASED MODELS

Bilal Kanso[1] and Omar Chebaro[2]

[1]Ecole Centrale Paris,Laboratoire de Mathématiques Appliquées aux Systèmes (MAS), Grande Voie des Vignes F-92295 Châtenay-Malabry, France
bilal.kanso.ecp@gmail.com
[2]ASCOLA (EMN-INRIA, LINA), École des Mines de Nantes
44307 Nantes, France
omar_chebaro@hotmail.com

*ABSTRACT*

*The contribution of this paper is threefold: first, it defines a framework for modelling component-based systems, as well as a formalization of integration rules to combine their behaviour. This is based on finite state machines (FSM). Second, it studies compositional conformance testing i.e. checking whether an implementation made of conforming components combined with integration operators is conform to its specification. Third, it shows the correctness of the global system can be obtained by testing the components involved into it towards the projection of the global specification on the specifications of the components. This result is useful to build adequate test purposes for testing components taking into account the system where they are plugged in.*

## 1. INTRODUCTION

Compositional testing [1, 2] becomes increasingly one of the most promising techniques for dealing with the state explosion problems in system testing. It consists in proving globally correctness of a system by checking locally correctness of its subsystems (or components). The main idea is to design, develop and validate each component independently in order to be widely used in a more large system, while a system is described recursively, at a higher level of abstraction, as interconnections of such components. The validation step is usually achieved using the conformance testing theory, which aims to checking the functional correctness of an implementation of a system with respect to its specification by means of experiments on the implementation. In the paradigm of automata-based compositional testing, component behaviours and their requirements are modelled as finite state machines (FSM) or labelled transition systems (LTS). The composition of components is commonly formalized as an operation taking components as well as the nature of their interactions to provide a description of a new more complex component.

In this paper, the models that we use to denote specifications of components are made of finite state machines extended to be able to encompass non-deterministic behaviours. The component models are structured and combined by means of a slight extension of the synchronous parallel operator [10, 27]. Although our framework is finite state machines rather than labelled transition systems, the conformance relation we use is a slight extension of the ioco relation [25] to our components called cioco. Our reason for choosing ioco to the detriment of relations issues from finite state machines such that trace inclusion or quasi-reduction [5, 6], is that ioco contrary to other relations, allows implementations not only to do what is specified, but also to do more than what is specified. This requirement of testing conformance has a fundamental role in testing practice [26].

Furthermore, in this paper, we will show that under some conditions, the conformance relation cioco is preserved over the synchronous parallel operator. From a practice point of view, this result means that making assumption that the synchronous parallel operator is well-implemented and preserves its specification, the composition of component implementations always conforms to the composition of their specifications, whether each implementation component is in conformance (according cioco) to its corresponding sub-specification.

This result finds a way to make system validation modular. Systems are tested, subsystems per subsystems, in a modular way, rather than "as a whole". Thus, explosion problems are less prone and debugging is greatly facilitated. However, it turns out that in practice, such a result is not enough. First, as the number of test case combinations is often huge, testing components in isolation would cause test cases that are important for the global system to be overlooked. Suppose a system S for computing student grade averages uses a calculator. Based on compositional testing result, to test S, we need to test the calculator in isolation. However, there is no way to ensure that important behaviours of the calculator involved in computing grade averages are covered by generated test cases (i.e. test cases only bringing into play addition and division for grades ranging from 0 to 20). Second, there is a need to test components in the context in which they are expected to be used. By way of example, the disaster of *Ariane* 5 in 1996 is caused by the absence of testing in context of a software component which was only tested for *Ariane* 4.

Following the projection approach in [2, 7], we equip our framework with a projection mechanism which enables us to easily retrieve all relevant information about subsystems. From global behaviours of a system, it helps capturing the behaviours of its sub-systems, that typically occur in the context of the whole system. Then, we will give in this paper, a new compositinality result that takes into account the behaviour of global system in which components are plugged in. This result helps to strengthen the quality of components by taking into account their involvement in the global system that encapsulates them. Furthermore, based on this result, specific test purposes can be generated to make component testing efficient by focusing on the way components are used in global systems.

The paper is structured as follows. Section 2 introduces the definition of our FSM-based components and the synchronous parallel operator. Section 3 presents our conformance testing theory for components. Section 4 shows the main limitation of the conformance testing techniques and studies the preservation of the conformance relation for the synchronous parallel operator. Section 5 introduces the compositinality result based on our projection mechanism. Section 6 recalls the related works. Section 7 concludes and presents the future works.

## 2. COMPONENTS AND SYSTEMS

A component is defined in our framework as a finite state machine in which the dependence between outputs and both current state and inputs is relaxed from a strict deterministic, to encompass also non determinism. A Finite State Machine (FSM) is a non-deterministic Mealy machine formally defined as follows:

**Definition 1 (Component) .** A component C is a 5-tuple (S, $s^0$, I, O, R) where:

- S is a finite set of states with the initial state $s^0 \in S$;
- I is a finite set of finite set of inputs: O is a finite set of outputs;
- $R \subseteq S \times I \times O \times S$ is the transition relation ;

**Comp(I, O)** denotes the set of components over the input-output signature (I,O).

In the following, for (s,i,o,s') in R, we simply write $s \xrightarrow{}_R s'$ and we represent a component in the standard way, by a directed edge-labelled graph where nodes represent states and edges represent transitions.

In our context, we are mainly interested by finite traces. Finite traces are finite sequences of couples (input|output) defined as follows:

**Definition 2 (Component finite traces).** Let $C = (S, s^0, I, O, R)$ be a component. The finite trace of a state s of C, noted **Trace$_C$(s)**, is the whole set of the finite input-output sequences $<i_0|o_0, \ldots, i_n|o_n>$ such that $\exists (s_0,\ldots, s_n, s_{n+1}) \in S^*$ of states where fir every j, $0 \leq j \leq n$, $s_j \xrightarrow{}_R s_{j+1}$ with $s_0 = s$.

Hence, Trace$_C$(C) is the set Trace$_c$($s^0$).

Several composition operators have been proposed in the literature to combine FSM-based components. The *sequential* composition (called also *cascade* composition) $C = \cdot (C_1, C_2)$ of two components $C_1$ and $C_2$ corresponds to a composition where both components $C_1$ and $C_2$ are interconnected side-by-side and the output of one is the input of the other [10, 27]. A reaction of C consists then of a reaction of both $C_1$ and $C_2$, where $C_1$ reacts first, produces its outputs, and then $C_2$ reacts. That is to say, when $C_1$ is triggered by an input i from the environment, $C_1$ executes i and the produced output is fed to $C_2$. The *double sequential* composition $\bowtie (C_1, C_2)$ is a composition in which the system can be triggered either by an input of $C_1$ and then feeds the output produced to $C_2$ or by an input of $C_2$ and then feeds the output produced to $C_1$. The *synchronous product* $\circledast (C_1, C_2)$ of two components $C_1$ and $C_2$ corresponds to a composition where both components $C_1$ and $C_2$ are executed independently or jointly, depending on the input. Hence, $C_1$ and $C_2$ are simultaneously executed when triggered by a joint input i that belongs to both inputs set of $C_1$ and $C_2$. The C*artesian product* [27] $\otimes (C_1, C_2)$ is a composition where both components are executed simultaneously when triggered by a pair of input values. The *concurrent composition* [27] $C = \oplus (C_1, C_2)$ of two components $C_1$ and $C_2$ corresponds to a composition where both components $C_1$ and $C_2$ are executed independently or jointly, depending on the input received from environment. It combines both choice and parallel compositions, in the sense $C_1$ and $C_2$ can be simultaneously executed when triggered by a pair of inputs $(i_1, i_2)$ ($i_1$ belongs to inputs set of $C_1$ and $i_2$ belongs to inputs set of $C_2$), ($i_1 \in I_1$ and $i_2 \in I_2$) or separately when triggered by an input i: if $i \in I_1$, then $C_1$ is executed and the reaction of C is that of $C_1$, otherwise $C_2$ is executed and the reaction of C is that of $C_2$. The *synchronous parallel* composition (called also *interleaving parallel* composition [10]) $C = \odot (C_1, C_2)$ of two components $C_1$ and $C_2$ is a composition in which both $C_1$ and $C_2$ are executed independently or jointly depending on the input, in such a way that each input action received by C from the environment consists exclusively of an input action of either $C_1$ or $C_2$ i.e. there is no common input action for $C_1$ and $C_2$. Indeed, when the global system receives an input which is supposed to be an input action of $C_1$, $C_1$ reacts by producing an output. If that output does not belong to the input set of $C_2$, the reaction of the global system consists only of the reaction of $C_1$. Otherwise, the output produced is directly fed to $C_2$ and the reaction of the global system consists of the reaction of both $C_1$ and $C_2$ (one falls into the same composition as the sequential composition). In the same manner, when the global system receives an input supposed to be an input action of $C_2$, $C_2$ reacts by producing an output. If that output does not belong to the input set of $C_1$, the reaction of the global system consists only of the reaction of $C_2$. Otherwise, the output produced is directly fed to $C_1$ and the reaction of the global system consists of the reaction of both $C_1$ and $C_2$. Further technical details about the different kinds of composition presented above can be found in textbooks such as [27].

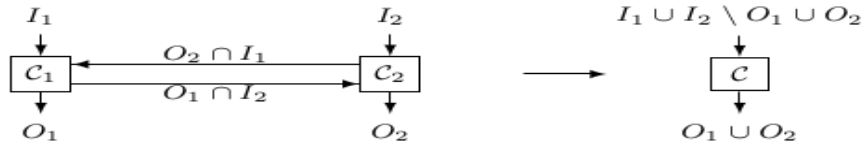

Fig. 1: Synchronous parallel composition $\odot$

It has been shown that the synchronous parallel operator is the most suitable and the more used operators to combine FSM-based components [10, 27]. Furthermore, it is generic enough to encompass some other integration operators. Indeed, this kind of composition can be seen as a general composition embodying both the synchronous and parallel aspects of composition. On one hand, it is synchronous in the sense that all common actions between $C_1$ and $C_2$ are synchronized. That means each output of $C_1$ that is fed as input of $C_2$ (i.e. $O_1 \cap I_2$) and each output of $C_2$ that is fed as input of $C_1$ ($O_2 \cap I_1$) are hidden (i.e. synchronized). They are not observable from the outside. On the other hand, it is parallel in the sense that both components $C_1$ and $C_2$ are considered autonomous: that is to say, a component may produce an output o regardless of whether o is specified as an input of the other component (see Fig. 1).

**Definition 3 (Synchronous parallel operator $\odot$).** Let $C_i = (S_i, s_i^0, I_i, O_i, R_i)$, for $i = 1, 2$, be two components such as $O_1 \cap I_2 \neq \emptyset$ and $O_2 \cap I_1 \neq \emptyset$. The **synchronous parallel operator** $\odot(C_1, C_2)$ is the 5-tuple $(S, s^0, I, O, R)$ where:

- $S = S_1 \times S_2$ and $s^0 = (s_1^0, s_2^0)$;
- $I = (I_1 \cup I_2) \setminus (O_1 \cup O_2)$ and $O = (O_1 \cup O_2)$;
- $R \subseteq S \times I \times O \times S$ is the minimal set satisfying the following inference rules:

  1. $C_1$ only reacts: $\dfrac{s_1 \xrightarrow{i|o}_{R_1} s_1',\ i \in I_1,\ o \notin I_2}{(s_1, s_2) \xrightarrow{i|o}_R (s_1', s_2)}$

  2. $C_2$ only reacts: $\dfrac{s_2 \xrightarrow{i|o}_{R_2} s_2',\ i \in I_2,\ o \notin I_1}{(s_1, s_2) \xrightarrow{i|o}_R (s_1, s_2')}$

  3. Both $C_1$ and $C_2$ react simultaneously, the output $o'$ of $C_1$ is synchronously consumed by $C_2$: $\dfrac{s_1 \xrightarrow{i|o'}_{R_1} s_1',\ s_2 \xrightarrow{o'|o}_{R_2} s_2'}{(s_1, s_2) \xrightarrow{i|o}_R (s_1', s_2')}$

  4. Both $C_1$ and $C_2$ react simultaneously, the output $o'$ of $C_2$ is synchronously consumed by $C_1$: $\dfrac{s_2 \xrightarrow{i|o'}_{R_2} s_2',\ s_1 \xrightarrow{o'|o}_{R_1} s_1'}{(s_1, s_2) \xrightarrow{i|o}_R (s_1', s_2')}$

**Note :** $\Box$ (resp. $\Box$) is similar to $\Box$ in which the set $I_1 \cap O_2 = \Box$ (resp. $O_2 = I_1$ and $O_1 = I_2$). $\Box$ and $\Box$ are particular cases of the concurrent composition.

Applying $\Box$ to basic components yield larger components that we will call *systems*. Then, given a set A of basic components, the set Sys(A) of systems over A is inductively defined as follows:

- For any component C ∈ A, C is in Sys(A);
- For any two components $C_1$, $C_2$ in Sys(A), ⊙ ($C_1$, $C_2$) is in Sys(A).

## 3. CONFORMANCE TESTING

Conformance testing [4] is a technique for checking the functional correctness of an implementation under test (*iut*) with respect to its specification (*spec*) by means of experiments on *iut*. It consists in deriving test cases algorithmically from a system specification, executing them on the real system and finally making sure that the latter behaves correctly by comparing its outputs with those required in the specification.

The notion of conformance is usually based on the comparison between the behaviour of a specification and an implementation using a conformance relation. The goal of this relation is to specify what the conformance of an implementation is to its specification. Several kinds of relations have been proposed in the literature. They differ mainly in both the formalism used to model system behaviour and the testing aspects considered. The original FSM-based conformance testing relation is defined as the testing equivalence of states whose goal is to determine the equivalence of two machines [3]. Two state machines are said to be *equivalent* if they produce exactly the same output sequence when offered the same input sequence. There is a list of other conformance relations that can be found in the literature. The definitions of these relations depend mainly on the underlying properties of the used finite state machines. Table 1 reviews some of them without going into details, for more detailed explanations, see [3, 5, 6].

| Relation | Informal definition | Properties |
| --- | --- | --- |
| **Equivalence** | Equality of traces set | Complete deterministic or complete non-deterministic |
| **Quasi Equivalence** | For each input sequence of *spec*, *spec* and *iut* produce the same output sequences | Deterministic or non-deterministic |
| **Reduction** | Trace inclusion | Complete non-deterministic |
| **Quasi reduction** | For each input sequence of *spec*, *iut* produces only output sequences of *spec* | Non-deterministic |

Table 1: Examples of conformance relations

It turns out that the conformance relations to test state equivalence are too strong, in practice, for conformance testing. There is a number of common assumptions (e.g. specification is strongly connected, minimized or complete) that are usually made in the literature to make test processes at all possible [3, 11, 15, 16]. Test generation algorithms based on them are also expensive in time and memory [17, 19, 18, 11 ,3], contrary to test cases generation techniques for inclusion relations (e.g. reduction and quasi reduction relations) [5, 6].

To cope with the weakness of FSM-based conformance relation, LTS model has been first appeared and some relations over it have been defined such as equivalence and pre-order relations relying on the observable behaviour notion [21,20]. However it turned rapidly out LTS formalism is so far to be applicable in testing practice due to the absence of a classification of actions into inputs and outputs [4, 14]. LTS then has been extended to Input-Output Labelled Transition System (IOLTS) in which there is a clear distinction between the input and outputs actions. For IOLTS model, several conformance relations were proposed such as the testing pre-order $\leq_{te}$, the refusal pre-order $\leq_{rf}$, ioconf and ioco [25, 26]. Among these relations, the relation ioco has received much attention by the community of formal testing because it has shown its suitability for conformance testing and especially automatic test derivation [25]. The reason is that the objective of conformance testing is mainly to check whether the implementation behaves as required by the specification i.e. to check if the implementation does what it should do. Hence, a conformance relation has to allow implementations not only to do what is specified, but also to do more than what is specified (for instance, when an annoyed user hits or kicks the coffee machine, or does other strange things that we are not usually considered in the specification). This requirement of testing conformance is well satisfied by ioco contrary to other relations [12, 20, 21] requiring testing behaviours that are not in the specification i.e. the implementation does not have the freedom to produce outputs for any input not considered in the specification.

The ioco relation that we will call here cioco (c for component) is formally redefined in terms of components as defined in Section 2. We make some modifications to the original definition of ioco to fit our component definition. That is, after each trace tr of a specification *spec*, instead of considering that the possible outputs of the corresponding implementation *iut* after executing tr on it is a subset of the possible outputs of *spec*, we consider that the corresponding implementation *iut*, after executing tr on it and then submitting any input i of the specification to it, does not produce outputs that are not allowed by *spec*.

**Definition 4.** (cioco) Let spec *be a specification over the signature* $(I, O)$ *and* iut *be an implementation over the same signature* $(I, O)$ *such as* iut *is input-enabled* [5]. iut *is said cioco* spec, *noted* iut cioco spec, *if and only if:*

$$\forall tr \in Trace(\text{spec}), \forall i \in I, Out(\text{iut after }(tr, i)) \subseteq Out(\text{spec after }(tr, i))$$

*where for any component* $\mathcal{C}$, *any finite trace* $tr = \langle i_0 | o_0, \ldots, i_n | o_n \rangle$ *of* $\mathcal{C}$, *and any input i of* $\mathcal{C}$:

$$Out(\mathcal{C} \text{ after }(tr, i)) = \{o \mid tr.\langle i|o\rangle \in Trace(\mathcal{C})\}$$

> **Note:** Commonly in conformance

nce testing, *iut* is assumed to be input-enabled i.e. it produces, at any state, answers for all possible inputs providing by the environment.

## 4. COMPONENT-BASED TESTING

As a matter of fact, the exponentially growing complexity and heterogeneity of today's systems give rise naturally to difficulties even the impossibility, in some cases, of using actual testing methods in practice. It turns out important aspects for software systems such as heterogeneity, decentralized and networked applications, etc. are not well-supported. This is especially due to the fact that testing techniques are limited to scalability of the complexity of actual software systems that are not only large but are also growing dramatically. As in a state-based components approach, compositional reasoning about system correctness is viewed as one of the most promising directions to bridge the gap between the increasing complexity of systems and actual testing method limits.

### 4.1. Approach

Component-based testing (or compositional testing) consists in testing communicating components that have been tested separately. It aims to guarantee the correctness of the integration of a set of components $op(C_1, \ldots, C_n)$ from the correctness of each components $C_i$ in isolation where op is the integration operator of interest. Thus, such a compositional testing theory provides a way to test the integrated system only by testing its sub-systems. As a consequence, there is no need to re-test its conformance correction. In our framework, the compositional testing problem is formally expressed as follows:

**Given ($iut_i$ cioco $spec_i$) for i = 1, 2, is it the case of ($iut_1$, $iut_2$) cioco ($spec_1$, $spec_2$)?**

Hence, once this question is positively answered, the correctness of the integrated system $\odot(iut_1, iut_2)$ is obtained from the correctness of the individual components $iut_1$ and $iut_2$. To test the integrated system, it is not necessary to consider it as a whole, but it is enough to consider its sub-systems and test them separately. Indeed, the contrapositive of this property is the following:

$$\neg(\odot(iut_1, iut_2) \text{ cioco } \odot(spec_1, spec_2)) \Rightarrow \exists i, i = 1, 2, \neg(iut_i \text{ cioco } spec_i)$$

Thus, by looking at this new property, we can easily see that non-correctness of $\odot(iut_1, iut_2)$ implies that at least one of $iut_1$ and $iut_2$ is incorrect. In other words, that means to test $\odot(iut_1, iut_2)$, it suffices to test $iut_1$ and $iut_2$ in isolation.

### 4.2. Illustration example

To illustrate our compositional testing, we consider two components of a coffee machine: a "money component" M that handles the inserted coins and "drink component" D that produces the drinks. Fig. 2 illustrates the architecture of these components.

**Note:** this example is inspired from the example presented in [1].

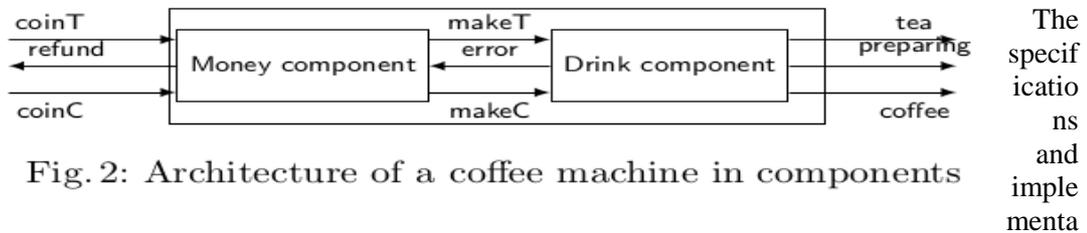

Fig. 2: Architecture of a coffee machine in components

The specifications and implementations of M and D are (see Fig. 3):

**Money component specification $spec_M$:** when it receives a coffee coin "coinC" (resp. a tea coin "coinT") from the user, it gives an order "makeC" (resp. "makeT") to the drink component D to make coffee (resp. tea).

**Drink component specification $spec_D$:** when it receives the order "makeC" (resp. "makeT") to make coffee (resp. tea) from the money component M, if there is nothing wrong during the drink preparation process, it directly delivers the coffee (resp. tea) to the user, or else it sends an error message to the money component in order to refund the user.

**Money component implementation $iut_M$:** it behaves as the money component specification $spec_M$, but in addition it does some extra functionalities, that is if an error occurs during the drink preparation process, it refunds the inserted coin to the user.

**Drink component implementation iut$_D$:** it behaves exactly as the drink component specification spec$_D$.

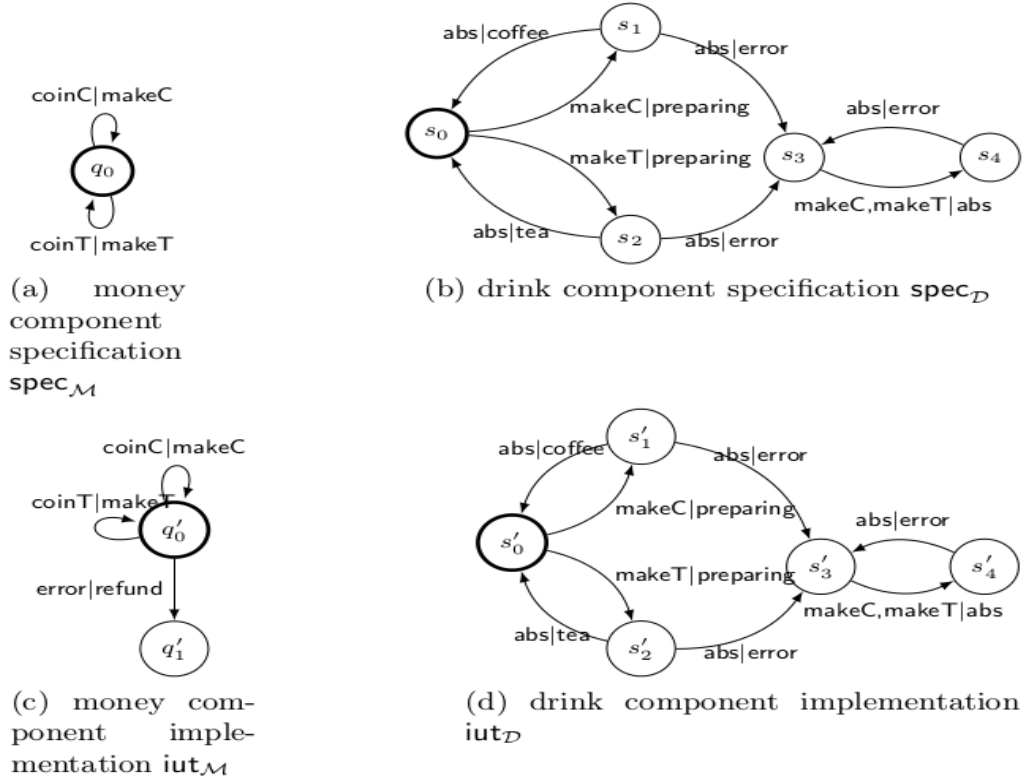

Fig. 3: Illustration of *cioco*'s compositionality

The components M and D may communicate separately (e.g. D may execute the transition labelled with abs|coffee while M does nothing) or jointly in synchronization (e.g. when M executes the transition labelled with coinC|makeC, M receives instantaneously the output makeC and then produces the output coffee). Then, such a composition of M and D is the synchronous parallel composition ∥ defined in Section 2.

**Note:** for the sake of readability, input completeness (implementations) are not depicted in Fig. 3c and Fig. 3d.

As far as the compositional testing is concerned, we have:

$$(iut_M \text{ cioco } spec_M) \text{ and } (iut_D \text{ cioco } spec_D)$$

Our goal is to know if this is enough to ensure whether the global implementation □(iut$_M$, iut$_D$) is in conformance w.r.t cioco to the global specification □(spec$_M$, spec$_D$). Hence, to test □(iut$_M$, iut$_D$), it suffices to test locally if (iut$_M$ cioco spec$_M$) and (iut$_D$ cioco spec$_D$). An answer to this question is given later in this paper.

### 4.3. Compositionality for synchronous parallel operator

We show here that the compositionality of cioco for synchronous parallel operator cannot be obtained without any assumptions made on both specifications and implementations. We first give an example that illustrates the assumptions required to obtain the compositionality of cioco w.r.t the synchronous parallel operator □. Figure 4 shows two implementation models *iut$_1$* and *iut$_2$* that have been tested to be cioco-correct according to their respective specification models

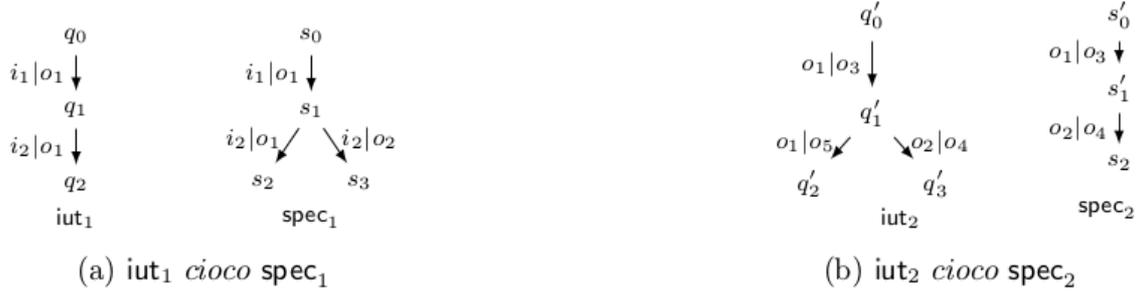

Fig. 4: Counterexample of compositionality

*spec$_1$* and *spec$_2$*. It is easy to see that (iut$_1$ cioco spec$_1$) and (iut$_2$ cioco spec$_2$).

Using the synchronous parallel operator □, the global implementation *iut* = (iut$_1$, iut$_2$) can do the trace $\langle i_1|o_3, i_2|o_5\rangle$. Thus, $o_5 \in \text{Out}(iut \text{ after } (\langle i_1|o_3\rangle, i_2))$ whereas the global specification *spec* = ⊙(spec$_1$, spec$_2$) can do the trace $\langle i_1|o_3\rangle$ in such a way $o_5 \notin \text{Out}(spec \text{ after } (\langle i_1|o_3\rangle, i_2))$. Hence, we can see that the global implementation *iut* does not conform to the global specification *spec* according to cioco.

This counterexample shows that ⊙ may give rise to a global implementation that does not conform to its global specification, even if the local implementations conform to their local specifications. The reason is that cioco does not put any constraint on the traces that are not specified in the specification. It allows implementations to do what they want with the unspecified states. Observe that if the specification specifies for any input what the allowed outputs are, then we do not have this problem. Hence, to cope with this problem, we assume that specifications are input-enabled as in [1]. That is to say, all states of a specification *spec* accept all input actions of *spec* (i.e. the transition relation R is total). Then, we have the following theorem for the compositionality for ⊙ :

**Theorem 1.** *Let* $\mathsf{spec}_1, \mathsf{iut}_1 \in \mathbf{Comp}(I_1, O_1)$ *and* $\mathsf{spec}_2, \mathsf{iut}_2 \in \mathbf{Comp}(I_2, O_2)$, *with* $I_1 \cap I_2 = O_1 \cap O_2 = \emptyset$ *such that* $\mathsf{spec}_1$ *and* $\mathsf{spec}_2$ *are input enabled. Then,* $(\mathsf{iut}_1 \; cioco \; \mathsf{spec}_1) \; and \; (\mathsf{iut}_2 \; cioco \; \mathsf{spec}_2) \Longrightarrow \odot(\mathsf{iut}_1, \mathsf{iut}_2) \; cioco \; \odot(\mathsf{spec}_1, \mathsf{spec}_2)$
*(See its proof in Appendix)*

Let us go back to the example presented in Subsection 4.2 where we have shown that:

$$(\mathsf{iut}_M \; cioco \; \mathsf{spec}_M) \text{ and } (\mathsf{iut}_D \; cioco \; \mathsf{spec}_D)$$

Here, the question is if:

$$\odot(\mathsf{iut}_M, \mathsf{iut}_D) \; cioco \; \odot(\mathsf{spec}_M, \mathsf{spec}_D)?$$

Our first attempt to answer this question is to check if the assumptions imposed in Theorem 1 are satisfied. Observe that neither *spec$_M$* nor *spec$_D$* are input-enabled. Hence, Theorem 1 fails to hold the compositinality of cioco for the components M and D. However, it is easy to see that the global implementation $\odot(\mathsf{iut}_M, \mathsf{iut}_D)$ can do the trace tr = <coinC|preparing, abs|coffee, coinC|preparing,abs|refund>. Thus:

$$\text{refund} \in \text{Out}(\text{iut after } (\text{<coinC|preparing, abs|coffee,coinC|preparing>, abs}))$$

whereas the global specification (spec$_M$, spec$_D$) can also do the trace <coinC|preparing, abs|coffee, coinC|preparing> in such a way:

$$\text{refund} \notin \text{Out}(\text{spec after } (\text{<coinC|preparing, abs|coffee, coinC|preparing>, abs}))$$

Hence, we can see that:

$$\neg((\mathsf{iut}_M, \mathsf{iut}_D) \; cioco \; (\mathsf{spec}_M, \mathsf{spec}_D))$$

## 5. TESTING IN CONTEXT

In Section 4, we have studied compositionality properties for cioco over the synchronous parallel operator $\odot$. We then proved that if single components of a system conform to their specifications, the whole system built over $\odot$ is in accordance with its specification, unless the specification model is input-enabled. However, it turns out that in practice, such a compositional approach is not enough. As an illustration, we consider an over simplified system

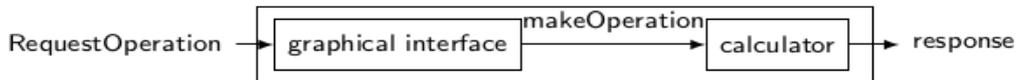

that computes grade averages. A typical design view of this system consists of two components:

1. An "user interface" ("controller") G that helps the user to make various operations on grades

2. A "calculator" C that receives operation commands from the user, performs the requested operation, and reports back to the user

According to the result obtained in Theorem 1, to test the grade average system, it suffices to test separately the calculator C and the controller G. However, testing the component C separately may lead to the consideration of test cases involving arithmetic operations which are irrelevant to computing student grade averages as subtraction, multiplication, square root, etc. This may cause test cases of interest to the system to be missed, i.e. test cases only bringing into play addition and division for grades ranging from 0 to 20.

In the following, we show how to improve significantly the result obtained in Theorem 1, by considering the global system in which components are plugged-in. We do so by defining projection mechanism. Such a projection mechanism is given in an accurate way by taking a behaviour p of the global system and keeping only the part of p being activated in the component that we want to test. This will allows us to generate more relevant unit test cases to test individual components.

### 5.1. Subsystem and projection

Given a system S= □(C, C'), we can inductively characterize the set of all basic (i.e. elementary) components, noted Sub(S), from which the global system S is built as follows:

– if C is a basic component, then Sub(S) = {C} ∪ Sub(C ');

– if C' is a basic component, then Sub(S) = {C'} ∪ Sub(C);

– otherwise (i.e. both C' and C' are not basic), then Sub(S) = Sub(C) ∪ Sub(C').

Projection techniques [2] are defined by pruning from any global behaviour p, all that does not concern the sub-system that we want to test. For any finite trace tr of a system S and a component C of S, we characterize the set of finite traces tr↓$_c$ of C involved in tr.

In the following definition, the notation:

$$s \xRightarrow{\eta.i|o}_R s'$$

means that the state s' is reachable from the state s, after the trace η following by i|o.

We then introduce the projection of a system, which we call *component in context*, on a one of its sub-systems.

**Definition 6 (Component in context).** Let $\mathcal{S}$ be a system over $(I, O)$ and $\mathcal{C} \in \mathbb{S}ub(\mathcal{S})$ be a component of $\mathcal{S}$ over $(I', O')$. The **component obtained by projecting $\mathcal{S}$ on $\mathcal{C}$**, noted $\mathcal{S}_{\downarrow c}$ is the 5-tuple $(S, s^0, I', O', R)$ defined by:

- $s^0 = \langle\rangle$
- $S$ is the whole set of finite traces defined as follows:
    - $s^0 = \{\langle\rangle\}$
    - $\forall j, 1 \leq j \leq n, s^j = \{tr'.\langle i|o\rangle \mid \exists tr' \in s^{j-1}, \exists (i,o) \in I' \times O', \exists tr \in Trace(\mathcal{S}) \text{ such that } tr'.\langle i|o\rangle \in tr_{\downarrow c}\}$
    
    Hence, $S = \bigcup_{j \geq 0} s^j$

- $R \subseteq S \times I' \times O' \times S$ is the minimal set satisfying the following rule:
    $\langle i_0|o_0, \ldots, i_m|o_m\rangle \xrightarrow{i|o}_R \langle i_0|o_0, \ldots, i_m|o_m, i|o\rangle$ iff: $\exists (i,o) \in I' \times O'$
    $\langle i_0|o_0, \ldots, i_m|o_m\rangle \in S$ and $tr \in Trace(\mathcal{S})$ such that $\langle i_0|o_0, \ldots, i_m|o_m, i|o\rangle \in tr_{\downarrow c}$

In this definition, S is the set of the states of the component in context. $s^0$ is the initial state of the component. Each state is represented by the unique sequence $\langle i_0|o_0, \ldots, i_n|o_n\rangle$ which leads to it from the initial state:

$$s^0 \xrightarrow{i_0|o_0} s_1 \xrightarrow{i_1|o_1} \ldots \xrightarrow{i_n|o_n} s_{n+1}$$

R gives, for each state s, and for each couple input-output i|o, the set of states that can be reached from s when the input i is submitted to the component.

**Note:** it is easy to see that the traces of the component $\mathcal{S}_{\downarrow c}$ obtained by projection is a subset of the traces of the component C itself.

### 5.2. Result

We here present our result of compositionality of testing. It consists in proving that the correctness of the integrated system is obtained from the correctness of the components given by projection of the global system on its components.

**Theorem 2.** Let $\text{spec}_1, \text{iut}_1 \in \mathbf{Comp}(I_1, O_1)$ and $\text{spec}_2, \text{iut}_2 \in \mathbf{Comp}(I_2, O_2)$, with $I_1 \cap I_2 = O_1 \cap O_2 = \emptyset$. Then, we have:

$$\left.\begin{array}{l}\text{iut}_1 \ cioco \ \odot(\text{spec}_1, \text{spec}_2)_{\downarrow \text{spec}_1} \\ \text{iut}_2 \ cioco \ \odot(\text{spec}_1, \text{spec}_2)_{\downarrow \text{spec}_2}\end{array}\right\} \Longrightarrow$$
$$\odot(\text{iut}_1, \text{iut}_2) \ cioco \ \odot(\text{spec}_1, \text{spec}_2)$$

(See its proof in Appendix)

Theorem 2 then provides a way to test the integrated system only by testing the projection of that system on its subsystems. As a consequence, to test the integrated system, it is not necessary to consider it as a whole, but it is enough to consider the projection of that system on its sub-systems (which may be done at different development steps and eventually developed by different teams) and test them separately. Comparing this result with our previous result presented in [13] or Tretmans's result [1], the new result does not require that the specifications are input-enabled. This last property is often hard to get in practice due to the fact that system input domains are usually too large. Let us replace, in the example presented in Subsection 4.2, the money specification by this presented in Figure 5. The projection $\Box(spec_M, spec_D){\downarrow}_{spec_M}$ of $\Box(spec_M, spec_D)$ on $spec_M$ is then the component $spec_M$ itself (Figure 5).

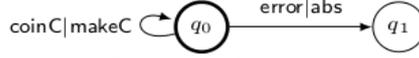

Fig. 5: New specification of the money component

According to Theorem 2, to test $\neg(iut_M$ cioco $spec_M)$, it is enough to test that:
$\neg(iut_M$ cioco $\Box(spec_M , spec_D){\downarrow}_{spec_M})$ or $\neg(iut_M$ cioco $\Box(spec_M , spec_D){\downarrow}_{spec_D})$

But, we know $\neg(iut_M$ cioco $\Box(spec_M , spec_D){\downarrow}_{spec_M})$ since after the trace <coinC|makeC> and for the input **error** of $spec_M$, the implementation $iut_M$ produces the output **refund** which is not allowed by the specification $spec_M$ (the only allowed output is **abs**). Hence, we can conclude that:
$\neg (\Box (iut_{M,} iut_D)$ cioco $\Box (spec_M, spec_D))$

## 6. RELATED WORK

Several compositional testing approaches have been proposed [1, 8, 13, 2, 10, 22, 7]. These approaches vary according to both formalism and integration operators. In [1], it has been proved that the conformance testing ioco based on labelled transition systems is only compositional w.r.t parallel composition when specifications and implementations are assumed input-enabled. In [8], it has been then shown that cspio (an adapted version of ioco to CSP formalism) is compositional not only for parallel composition but also for other CSP's composition operators by assuming input completeness of the specification in the same alphabet of the implementation. The authors of [13] use co-algebra theory to obtain generic result of compositional testing. They propose to extend component-based testing approach [1] to co-algebraic components [24]. In [2, 10, 22, 7], the authors address differently the compositional testing problem from [1,8, 13]. In [2], the authors work with input-output symbolic transition systems (IOSTS) and propose to test each component of a system in isolation by generating accurate test purposes for them from the global system specification and assuming that the specification of every component in the system is available. This allowed them to test the global system by selecting behaviours of basic components that are typically activated in the system, and then re-enforce unitary testing w.r.t those behaviours. However, there is no testing compositinality result. In [10], the authors study how to design a component when combined with a known part of the system, called the context, has to satisfy a given overall specification in the context of finite state machine. In [22], the authors extend the so-called assume-guarantee reasoning [9] used in model checking areas as a means to cope with the state explosion problem of compositional testing. They then proposed to test each component of a system separately, while taking into account assumptions about the context of the component. They use the input-output labelled transition systems as behavioural models of components and the parallel composition to compose components. The conformance relation used in this approach is the ioco relation. The underlying idea behind this approach is to check that, given a assumption A about the environment in which the components are supposed to operate, such that ($iut_2$ ioco A) and (($iut_1$ || A) ioco spec) then (($iut_1$ || $iut_2$) ioco spec). The authors showed that this property

holds if the assumption A is input-enabled. Finally, in [7], the authors propose to extend [2] in order to be able to generate test purposes for co-algebraic components [13, 24].

## 7. CONCLUSION

This paper defines a framework for modelling and testing component-based systems. On the first hand, we have proposed a FSM-based framework for modelling systems and we have defined the synchronous parallel operator for combining component behaviour. On the second hand, we have proved two compositional testing result. The first one, assuming input completeness of the specification model in the same alphabet of the implementation model, that compositionality holds for the synchronous parallel operator. The second one shows, using projection mechanism, that compositionality naturally holds for the synchronous parallel operator.

For future work, we will be interested in proposing an approach to generate adequate test purposes automatically that focus mainly on component behaviours which are activated in the global system. The underlying idea is to build for a trace tr of the global system a finite computation tree for the component involved in tr. Then, using the algorithm proposed in [2] to generate correct test cases for individual components.

**Authors**

Bilal Kanso holds a teaching and research position since September 2012 at Paris 12 University. He was a post-doctoral fellow at the Computer Science department of Supélec France. He participated in the development of a technique for adding temporal operators to the Object Constraint Language (OCL) in order to express temporal properties of the object-oriented programs, with verification issues in mind. He obtained his PHD degree from the Ecole Centrale Paris in November 2011. His dissertation deals with formal methods and software testing using both conformance testing and coalgebra theories. His research also includes model checking and its application to object-oriented programming and compositional verification and testing methods.

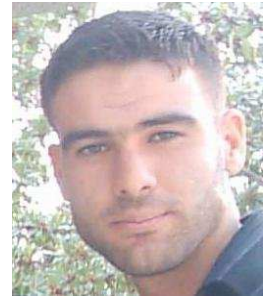

Omar Chebaro holds a postdoc position since January 2012 in the ASCOLA team, a joint team of EMNante and INRIA. He obtained his PHD degree from the university of Franche-Comté in December 2011. His dissertation deals with formal methods and software verification using static analysis, program transformation and dynamic analysis. His research also includes concrete and symbolic test generation, label coverage and compositional verification.

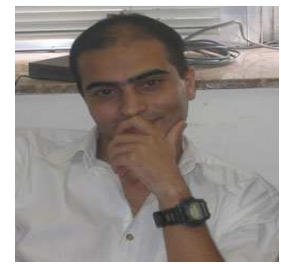

# Appendix

**Proof Theorem 1**

In order to prove Theorem 1, we need to prove the two following lemmas:

**Lemma 1.** *Let $C_i = (S_i, s_i^0, I_i, O_i, R_i) \in \mathbf{Comp}(I_i, O_i)$ with $i = 1, 2$. Let $C = (S, s^0, I, O, R) = \odot(C_1, C_2)$, then:*
$(s_1^0, s_1^0) \stackrel{\sigma}{\Longrightarrow}_R (s_1, s_2) \Leftrightarrow$
$\exists tr_1 \in \sigma_{\downarrow C_1}, \sigma_2 \in tr_{\downarrow C_2} : s_1^0 \stackrel{tr_1}{\Longrightarrow}_{R_1} s_1 \text{ and } s_2^0 \stackrel{tr_2}{\Longrightarrow}_{R_2} s_2$

*Proof.*
**Only if:** *proof by induction on the structure of $tr$. Let $tr \in Trace(C)$*

**Basic step:** *if $(s_1^0, s_1^0) \stackrel{\epsilon}{\Longrightarrow}_R (s_1, s_2)$, then $\exists s_1' \in S_1, s_2' \in S_2 : s_1^0 \stackrel{\epsilon}{\Longrightarrow}_{R_1} s_1'$ and $s_2^0 \stackrel{\epsilon}{\Longrightarrow}_{R_2} s_2'$ trivially holds.*

**Induction step:** *we make the assumption that the lemma holds for $tr = \sigma.\langle i|o\rangle \in Trace(C)$. Then, $(s_1^0, s_1^0) \stackrel{\sigma.\langle i|o\rangle}{\Longrightarrow}_R (s_1, s_2)$. By definition of $\stackrel{tr}{\Longrightarrow}$, there exist $(r_1, r_2) \in S$ such that $(s_1^0, s_1^0) \stackrel{\sigma}{\Longrightarrow}_R (r_1, r_2) \stackrel{\langle i|o\rangle}{\Longrightarrow}_R (s_1, s_2)$. Now, by hypothesis, we can conclude that: $\exists \sigma_1 \in \sigma_{\downarrow C_1}, \sigma_2 \in \sigma_{\downarrow C_2} : s_1^0 \stackrel{\sigma_1}{\Longrightarrow}_{R_1} r_1, s_2^0 \stackrel{\sigma_2}{\Longrightarrow}_{R_2} r_2$ and $(r_1, r_2) \stackrel{\langle i|o\rangle}{\Longrightarrow}_R (s_1, s_2)$. We identify four cases:*

1. $\mathbf{i \in I_1 \setminus O_2, o \notin I_2}$: *the projection of $\langle i|o\rangle$ on $C_1$ and $C_2$ are:*

   $$\begin{cases} \langle i|o\rangle_{\downarrow C_1} = \{\langle i|o\rangle\} \\ \langle i|o\rangle_{\downarrow C_2} = \epsilon \end{cases}$$

   *Then, we can conclude that there exists $tr_1 = \sigma_1.\langle i|o\rangle \in tr_{\downarrow C_1}$ and $tr_2 = \sigma_2.\epsilon \in tr_{\downarrow C_2}$ such that $s_1^0 \stackrel{tr_1}{\Longrightarrow}_{R_1} s_1'$ and $s_2^0 \stackrel{tr_2}{\Longrightarrow}_{R_2} s_2'$*

2. $\mathbf{i \in I_2 \setminus O_1, o \notin I_1}$: *the projection of $\langle i|o\rangle$ on $C_1$ and $C_2$ are:*

   $$\begin{cases} \langle i|o\rangle_{\downarrow C_1} = \epsilon \\ \langle i|o\rangle_{\downarrow C_2} = \{\langle i|o\rangle\} \end{cases}$$

   *Then, we can conclude that there exists $tr_1 = \sigma_1.\epsilon \in tr_{\downarrow C_1}$ and $tr_2 = \sigma_2.\langle i|o\rangle \in tr_{\downarrow C_2}$ such that $s_1^0 \stackrel{tr_1}{\Longrightarrow}_{R_1} s_1'$ and $s_2^0 \stackrel{tr_2}{\Longrightarrow}_{R_2} s_2'$*

3. $\mathbf{i \in I_1, o \in I_2}$: *the projection of $\langle i|o\rangle$ on $C_1$ and $C_2$ are:*

   $$\begin{cases} \langle i|o\rangle_{\downarrow C_1} = \{\langle i|o'\rangle \mid \exists s_1, s_1' \text{ s.t } s_1 \stackrel{\eta.i|o'}{\Longrightarrow}_{R_1} s_1' \text{ and} \\ \qquad \exists s_2, s_2' \text{ s.t } s_2 \stackrel{\eta.o|o'}{\Longrightarrow}_{R_2} s_2'\} \\ \langle i|o\rangle_{\downarrow C_2} = \{\langle o'|o\rangle \mid \exists s_1, s_1' \text{ s.t } s_1 \stackrel{\eta.i|o'}{\Longrightarrow}_{R_1} s_1' \text{ and} \\ \qquad \exists s_2, s_2' \text{ s.t } s_2 \stackrel{\eta.o|o'}{\Longrightarrow}_{R_2} s_2'\} \end{cases}$$

Then, we can conclude that there exists $tr_1 = \sigma_1.\langle i|o'\rangle \in tr_{\downarrow_{\mathcal{C}_1}}$ and $tr_2 = \sigma_2.\langle o'|o\rangle \in tr_{\downarrow_{\mathcal{C}_2}}$ such that $s_1^0 \stackrel{tr_1}{\Longrightarrow}_{R_1} s_1'$ and $s_2^0 \stackrel{tr_2}{\Longrightarrow}_{R_2} s_2'$

4. $\mathbf{i \in I_2, o \in O_1}$: the projection of $\langle i|o\rangle$ on $\mathcal{C}_1$ and $\mathcal{C}_2$ are:

$$\begin{cases} \langle i|o\rangle_{\downarrow_{\mathcal{C}_1}} = \{\langle o'|o\rangle \mid \exists s_1, s_1' \text{ s.t } s_1 \stackrel{\eta.i|o'}{\Longrightarrow}_{R_1} s_1' \text{ and} \\ \qquad \exists s_2, s_2' \text{ s.t } s_2 \stackrel{\eta.o|o'}{\Longrightarrow}_{R_2} s_2'\} \\ \langle i|o\rangle_{\downarrow_{\mathcal{C}_2}} = \{\langle i|o'\rangle \mid \exists s_1, s_1' \text{ s.t } s_1 \stackrel{\eta.i|o'}{\Longrightarrow}_{R_1} s_1' \text{ and} \\ \qquad \exists s_2, s_2' \text{ s.t } s_2 \stackrel{\eta.o|o'}{\Longrightarrow}_{R_2} s_2'\} \end{cases}$$

Then, we can conclude that there exists $tr_1 = \sigma_1.\langle o'|o\rangle \in tr_{\downarrow_{\mathcal{C}_1}}$ and $tr_2 = \sigma_2.\langle i|o'\rangle \in tr_{\downarrow_{\mathcal{C}_2}}$ such that $s_1^0 \stackrel{tr_1}{\Longrightarrow}_{R_1} s_1'$ and $s_2^0 \stackrel{tr_2}{\Longrightarrow}_{R_2} s_2'$

The **if** part can be proved in the same manner.

**Lemma 2.** *Consider two components $\mathcal{C}_1$ and $\mathcal{C}_2$, then:*

1. $Trace(\mathcal{C}_1) \subseteq Trace(\mathcal{C}_2)$ *implies* $(\mathcal{C}_1 \text{ cioco } \mathcal{C}_2)$
2. *If $\mathcal{C}_2$ is **input-enabled**, then $(\mathcal{C}_1 \text{ cioco } \mathcal{C}_2)$ implies $Trace(\mathcal{C}_1) \subseteq Trace(\mathcal{C}_2)$.*

*Proof.*

1. Let $tr = \langle i_1|o_1, \ldots, i_n|o_n\rangle$ be a finite trace of $\mathcal{C}_2$, $i$ an input of $\mathcal{C}_2$ and $o \in Out(\mathcal{C}_1 \text{ after } (tr, i))$ and let us prove that $o \in Out(\mathcal{C}_2 \text{ after } (tr, i))$.
   $o \in Out(\mathcal{C}_1 \text{ after } (tr, i))$ implies $tr' = tr.\langle i|o\rangle = \langle i_1|o_1, i_2|o_2, \ldots, i_n|o_n, i|o\rangle \in Trace(\mathcal{C}_1)$. Since $Trace(\mathcal{C}_1) \subseteq Trace(\mathcal{C}_2)$, then $tr' \in Trace(\mathcal{C}_2)$. Thus, $o \in Out(\mathcal{C}_2 \text{ after } (tr, i))$, and consequently,

   $$Out(\mathcal{C}_1 \text{ after } (tr, i)) \subseteq Out(\mathcal{C}_2 \text{ after } (tr, i))$$

   The result then follows from the definition of cioco.

2. By induction on the structure of a trace $tr$ of $\mathcal{C}_1$. Let $tr = \langle i_1|o_1, \ldots, i_n|o_n\rangle \in Trace(\mathcal{C}_1)$.
   - **Basic Step:** $tr = \langle\rangle$ is empty trace.
     $tr = \langle\rangle \in Trace(\mathcal{C}_2)$ trivially holds.
   - **Induction Step:** Let us write $tr$ as concatenation of two finite traces as follows:
     $$tr = \langle i_1|o_1, i_2|o_2, \ldots, i_{n-1}|o_{n-1}\rangle \cdot \langle i_n|o_n\rangle$$
     $tr \in Trace(\mathcal{C}_1)$ implies $o_n \in Out(\mathcal{C}_1 \text{ after } (\langle i_1|o_1, \ldots, i_{n-1}|o_{n-1}\rangle, i_n))$. Since $\mathcal{C}_2$ is input-enabled, $i_n$ is inevitably an input of $\mathcal{C}_2$ at any state $s$. By induction hypothesis, we have:
     $$\langle i_1|o_1, \ldots, i_{n-1}|o_{n-1}\rangle \in Trace(\mathcal{C}_2)$$

and
$$o_n \in Out(\mathcal{C}_1 \text{ after } (\langle i_1|o_1,\ldots,i_{n-1}|o_{n-1}\rangle, i_n))$$

Then $o_n \in Out(\mathcal{C}_2 \text{ after } (\langle i_1|o_1,\ldots,i_{n-1}|o_{n-1}\rangle, i_n))$ because ($\mathcal{C}_1$ cioco $\mathcal{C}_2$). Thus $\langle i_1|o_1,\ldots,i_{n-1}|o_{n-1}, i_n|o_n\rangle \in Trace(\mathcal{C}_2)$. Consequently, $Trace(\mathcal{C}_1) \subseteq Trace(\mathcal{C}_2)$.

*Proof.* Let us now prove Theorem 1. According to Lemma 2, we have to prove:
$$Trace(\odot(\text{iut}_1, \text{iut}_2)) \subseteq Trace(\odot(\text{spec}_1, \text{spec}_2))$$

Let $s_{i1}^0$ and $s_{i2}^0$ be the initial states of $\text{iut}_1$ and $\text{iut}_2$ resp. and $s_{s1}^0$ and $s_{s2}^0$ be the initial states of $\text{spec}_1$ and $\text{spec}_2$ resp. Then, we have:

| | |
|---|---|
| $tr \in Trace(\odot(\text{iut}_1, \text{iut}_2))$ | Definition of Traces |
| $\Rightarrow (s_{i1}^0, s_{i2}^0) \stackrel{tr}{\Longrightarrow}$ | Lemma 1 |
| $\Rightarrow s_{i1}^0 \stackrel{tr_1}{\Longrightarrow} \wedge s_{i2}^0 \stackrel{tr_2}{\Longrightarrow}$ | Premise |
| $\Rightarrow s_{s1}^0 \stackrel{tr_1}{\Longrightarrow} \wedge s_{s2}^0 \stackrel{tr_2}{\Longrightarrow}$ | Lemma 1 |
| $\Rightarrow (s_{s1}^0, s_{s2}^0) \stackrel{tr}{\Longrightarrow}$ | Definition of Traces |
| $\Rightarrow tr \in Trace(\odot(\text{spec}_1, \text{spec}_2))$ | |

**Proof Theorem 2**

*Proof.* Let $\text{iut} = \odot(\text{iut}_1, \text{iut}_2)$ and $\text{spec} = \odot(\text{spec}_1, \text{spec}_2)$. Let us assume that
$$\text{iut}_1 \; cioco \; \text{spec}_{\downarrow\text{spec}_1} \quad \text{and} \quad \text{iut}_2 \; cioco \; \text{spec}_{\downarrow\text{spec}_2} \quad (\text{Hyp})$$

and then prove that ($\text{iut} \; cioco \; \text{spec}$).
Let $tr = \langle i_1|o_1,\ldots,i_n|o_n\rangle \in Trace(\text{iut}) \cap Trace(\text{spec})$, $i \in (I_1 \cup I_2) \setminus (O_1 \cup O_2)$ and $o \in O_1 \cup O_2$ such that $o \in Out(\text{iut after } (tr, i))$ and let us prove that $o \in Out(\text{spec after } (tr, i))$.

We have
$$tr = \langle i_1|o_1,\ldots,i_n|o_n\rangle \in Trace(\text{iut})$$

According to Lemma 1, there exist two traces $tr_1 \in tr_{\downarrow\text{iut}_1} \subseteq Trace(\text{iut}_1)$ and $tr_2 \in tr_{\downarrow\text{iut}_2} \subseteq Trace(\text{iut}_2)$ that are respectively the traces involved in $\text{iut}_1$ and $\text{iut}_2$ to obtain $tr$. We also know by the projection definition (see Definition **??**) and (Hyp) that $tr_1 \in tr_{\downarrow\text{spec}_1} \subseteq Trace(\text{spec}_{\downarrow\text{spec}_1})$ and $tr_2 \in tr_{\downarrow\text{spec}_2} \subseteq Trace(\text{spec}_{\downarrow\text{spec}_2})$. Hence, we can conclude:

$$tr_1 \in Trace(\text{iut}_1) \cap Trace(\text{spec}_{\downarrow\text{spec}_1}) \text{ and } tr_2 \in Trace(\text{iut}_2) \cap Trace(\text{spec}_{\downarrow\text{spec}_2}) \quad (2)$$

Now, we have $o \in Out(\text{iut after } (tr, i))$, we identify the following cases:

1. $i \in \mathbf{I_1}, o \notin \mathbf{O_2}$: according to the projection definition, we have $tr_1.\langle i|o\rangle \in Trace(\mathsf{iut}_1)$. That implies $o \in Out(\mathsf{iut}_1 \text{ after } (tr_1, i))$. We know $\mathsf{iut}_1 \; cioco \; \mathsf{spec}_{\downarrow\mathsf{spec}_1}$, then $o \in Out(\mathsf{spec}_{\downarrow\mathsf{spec}_1} \text{ after } (tr_1, i))$. Hence
$$tr_1.\langle i|o\rangle \in Trace(\mathsf{spec}_{\downarrow\mathsf{spec}_1}) \subseteq Trace(\mathsf{spec}_1)$$
By the definition of $\odot$, we can conclude that $tr.\langle i|o\rangle \in Trace(\mathsf{spec})$. Consequently, $o \in Out(\mathsf{spec} \text{ after } (tr, i))$.

2. $i \in \mathbf{I_1}, o \in \mathbf{O_2}$: according to the projection definition, there exists $o' \in O_1$ such that $tr_1.\langle i|o'\rangle \in Trace(\mathsf{iut}_1)$ and $tr_2.\langle o'|o\rangle \in Trace(\mathsf{iut}_2)$. That means $o' \in Out(\mathsf{iut}_1 \text{ after } (tr_1, i))$ and $o \in Out(\mathsf{spec}_2 \text{ after } (tr_2, o'))$. According to (Hyp), we have:
$$o' \in Out(\mathsf{spec}_{\downarrow\mathsf{spec}_1} \text{ after } (tr_1, i)) \text{ and } o \in Out(\mathsf{spec}_{\downarrow\mathsf{spec}_2} \text{ after } (tr_2, o'))$$
Hence
$$tr_1.\langle i|o'\rangle \in Trace(\mathsf{spec}_{\downarrow\mathsf{spec}_1}) \subseteq Trace(\mathsf{spec}_1)$$
and
$$tr_2.\langle o'|o\rangle \in Trace(\mathsf{spec}_{\downarrow\mathsf{spec}_2}) \subseteq Trace(\mathsf{spec}_2)$$
By the definition of $\odot$, we can conclude that $tr.\langle i|o\rangle \in Trace(\mathsf{spec})$. Consequently, $o \in Out(\mathsf{spec} \text{ after } (tr, i))$.

3. $i \in \mathbf{I_2}, o \in \mathbf{O_1}$: according to the projection definition, there exists $o' \in O_2$ such that $tr_1.\langle o'|o\rangle \in Trace(\mathsf{iut}_1)$ and $tr_2.\langle i|o'\rangle \in Trace(\mathsf{iut}_2)$. That means $o \in Out(\mathsf{iut}_1 \text{ after } (tr_1, o'))$ and $o' \in Out(\mathsf{spec}_2 \text{ after } (tr_2, i))$. According to (Hyp), we have: $o \in Out(\mathsf{spec}_{\downarrow\mathsf{spec}_1} \text{ after } (tr_1, o'))$ and $o' \in Out(\mathsf{spec}_{\downarrow\mathsf{spec}_2} \text{ after } (tr_2, i))$
Hence
$$tr_1.\langle o'|o\rangle \in Trace(\mathsf{spec}_{\downarrow\mathsf{spec}_1}) \subseteq Trace(\mathsf{spec}_1)$$
and
$$tr_2.\langle i|o'\rangle \in Trace(\mathsf{spec}_{\downarrow\mathsf{spec}_2}) \subseteq Trace(\mathsf{spec}_2)$$
By the definition of $\odot$, we can conclude that $tr.\langle i|o\rangle \in Trace(\mathsf{spec})$. Consequently, $o \in Out(\mathsf{spec} \text{ after } (tr, i))$.

4. $i \in \mathbf{I_2}, o \notin \mathbf{O_1}$: according to the projection definition, we have $tr_2.\langle i|o\rangle \in Trace(\mathsf{iut}_2)$. That implies $o \in Out(\mathsf{iut}_2 \text{ after } (tr_2, i))$. We know $\mathsf{iut}_2 \; cioco \; \mathsf{spec}_{\downarrow\mathsf{spec}_2}$, then $o \in Out(\mathsf{spec}_{\downarrow\mathsf{spec}_2} \text{ after } (tr_2, i))$. Hence
$$tr_2.\langle i|o\rangle \in Trace(\mathsf{spec}_{\downarrow\mathsf{spec}_2}) \subseteq Trace(\mathsf{spec}_2)$$
By the definition of $\odot$, we can conclude that $tr.\langle i|o\rangle \in Trace(\mathsf{spec})$. Consequently, $o \in Out(\mathsf{spec} \text{ after } (tr, i))$.